\documentclass[final,5p,times,twocolumn]{elsarticle}

\usepackage{graphicx}
\usepackage{epsfig}
\usepackage{amssymb}
\usepackage{amsmath}
\usepackage{amsthm}
\usepackage{lineno}

\journal{Nuclear Instruments and Methods A}

\begin{document}

\begin{frontmatter}

%% Title, authors and addresses

\title{A UV Sensitive Integrated Micromegas with Timepix Readout}

\author[A]{Joost Melai}
\author[B]{Amos Breskin} 
\author[B]{Marco Cortesi}
\author[C]{Yevgen Bilevych}
\author[C]{Martin Fransen}
\author[C]{Harry van der Graaf}
\author[C]{Jan Visschers}
\author[C]{Victor Blanco Carballo}
\author[A]{Cora Salm\corref{cor1}}
\ead{c.salm@utwente.nl}
% \ead[url]{http://sc.el.utwente.nl/persons/?id=164}
\author[A]{Jurriaan Schmitz}
\address[A]{MESA$^+$ Institute for nanotechnology, University of Twente, Enschede, the Netherlands}
\address[B]{Weizmann Institute of Science, Rehovot, Israel}
\address[C]{NIKHEF, Amsterdam, the Netherlands}

\cortext[cor1]{Corresponding author, Telephone +31.53.489.2648}

\begin{abstract} % abstract max 10 lines
This article presents a detector system consisting of three components, a CMOS imaging array, a gaseous-detector structure with a Micromegas layout and a UV-photon sensitive CsI reflective photocathode. All three elements have been monolithically integrated using simple post-processing steps. The Micromegas structure and the CMOS imaging chip are not impacted by the CsI deposition.  The detector operated reliably in He/isobutane mixtures and attained charge gains with single photons up to a level of $6\cdot10^4$.
The Timepix CMOS array permitted high resolution imaging of single UV-photons. The system has an MTF50 of \mbox{0.4 lp/pixel} which corresponds to app. \mbox{7 lp/mm}.

\end{abstract}

\begin{keyword}
Gaseous radiation detector \sep Micromegas \sep UV photon detection \sep InGrid \sep CsI photocathode \sep CMOS post-processing
%% PACS codes here, in the form: \PACS code \sep code
%% MSC codes here, in the form: \MSC code \sep code
%% or \MSC[2008] code \sep code (2000 is the default)
\end{keyword}

\end{frontmatter}

%\linenumbers

% main text

\section{Introduction}
We present a novel integrated gaseous UV-photon detector that is made by post-processing a charge-sensitive pixilated CMOS imaging detector. It is based on the InGrid technology \cite{VictorEDL} which uses microtechnology techniques to construct a Micromegas detector structure on top of the readout chip. This method allows aligning the different components of the detector with the anode plane. The pillars are placed at the intersections of four adjacent pixels; the circular holes in the grid are centered onto the input pad of each pixel. The hole pitch equals that of the readout matrix, here 55~$\mu$m. The hole diameter can be varied, typical values are 20--30 $\mu$m. This affects the optical transparency of the grid, for the range mentioned it is \mbox{10--23\%}.
The photosensitivity is obtained by coating the grid with a reflective CsI photocathode \cite{CsIBreskin}. CsI is a very robust photocathode that has been used in combination with many gaseous photon detectors \cite{Chechik}.  In \cite{GioCsI} a Micromegas device with a single anode is coupled to a CsI reflective photocathode. A CMOS imaging chip coupled to a non-integrated gaseous electron multiplier (GEM) with CsI photocathode was described in \cite{Bellazzini}. This work presents the first fully monolithic embodiment of such a UV-photon imaging detector.

\section{Experimental}
\begin{figure}[hbt]
\centering
\includegraphics[width=0.35\textwidth,keepaspectratio]{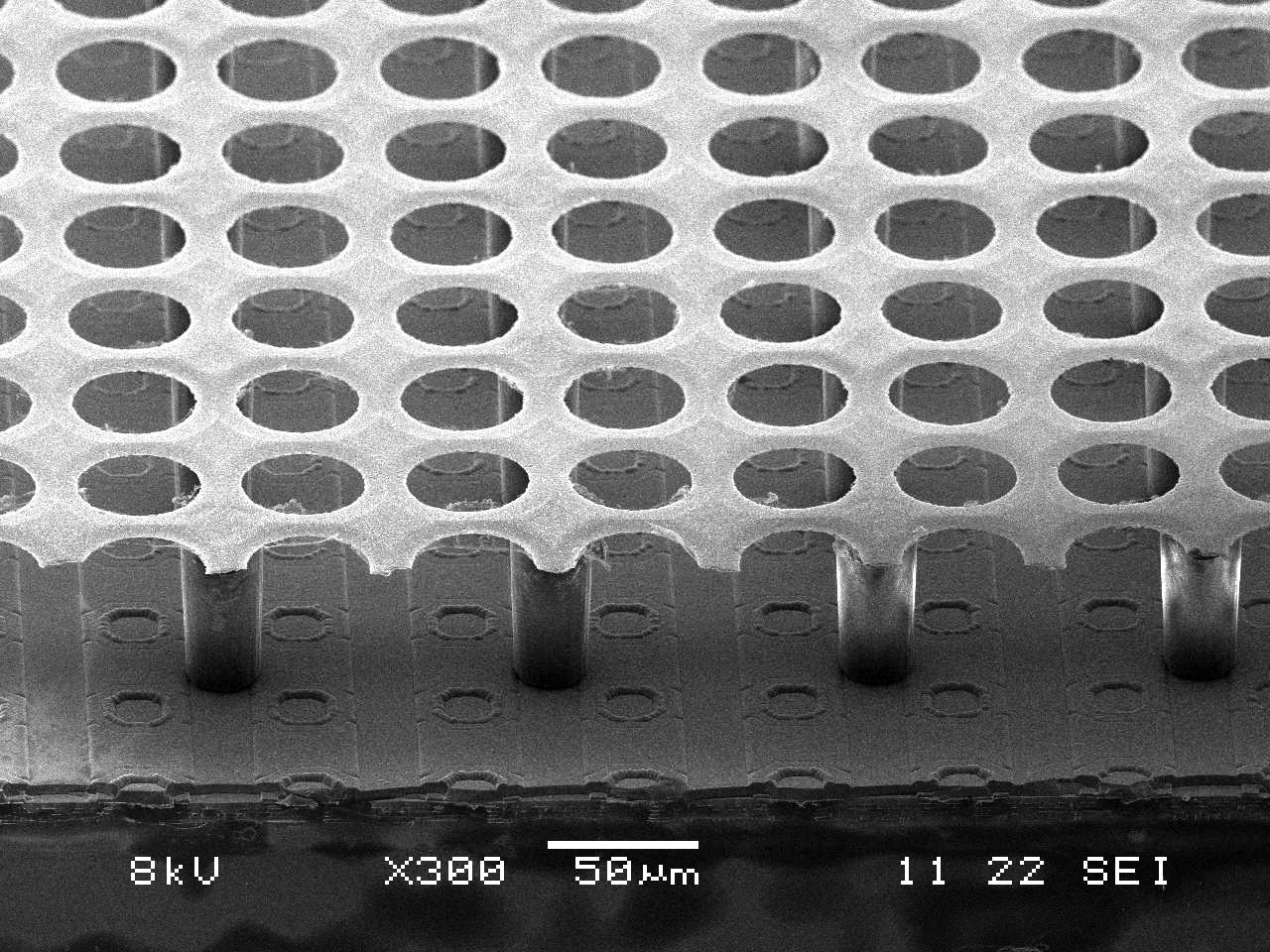}
\caption{SEM image of a complete detector with Timepix substrate, highly resistive spark protection layer and InGrid structure with isolating pillars supporting a perforated Al grid.}
\label{fig:InGridSEM}
\end{figure}

The detectors are comprised of three elements; the imaging chip, the integrated Micromegas grid and a reflective photocathode. Fig. \ref{fig:InGridSEM} shows the full detector assembly.
The Micromegas structure, the photocathode and some of the experimental methods discussed below are similar to our earlier work reported in \cite{IWORID}.

The Timepix chip used here \cite{timepix} is a variant of the Medipix2 chip \cite{medipix}. It has been designed for the Medipix2 collaboration by CERN, within the EUDET framework. The Timepix chip contains a matrix of $256 \times 256$ square pixels with a size of \mbox{55 $\mu$m}. The total imaging area is \mbox{$\sim 14 \times 14 \textrm{ mm}^2$}. Each of the pixels has a large metal input pad connected to an input capacitor. The collected charge is read out with a charge sensitive pre-amplifier capable of detecting charges of less than \mbox{1000 e$^-$}. The charge is compared with a threshold value which can be changed per pixel. There are different readout modes. Just as Medipix2 the chip can be used to count the number of hits above a threshold during a defined shutter period. In Timepix-mode the arrival time of the first hit above threshold is recorded per pixel and in TOT-mode (Time Over Threshold) the \mbox{14 bit} counter is incremented as long as the signal remains above threshold. Because the discharge current of the input capacitor is kept constant at a known value the TOT readout can be used to measure input charge.

\subsection{Chip post-processing}
The detector is built directly on the surface of the imaging chip as described in detail in \cite{VictorEDL}. The present devices have been built on single chips but the process can also be performed on clusters of several chips (for instance a square of 3 by 3 chips) or on entire wafers.

The first step is to cover the anode surface of the pixel matrix with a highly resistive layer, in our case this is a 7 $\mu$m Si-rich Si$_x$N$_y$ layer deposited by plasma-enhanced CVD. This layer protects the chip by quenching discharges that may occur in the gas during detector operation.

In a next phase the InGrid (Integrated Grid) structure is built. First a photoimagable SU\nobreakdash-8 polymer layer is spin cast onto the substrate. This layer is baked to render it solid. A \mbox{1 $\mu$m} thick Al layer is sputtered onto the SU\nobreakdash-8. To form the perforated grid the metal is patterned using standard lithography and wet etch techniques. The SU-8 layer, which has been imaged earlier, is then developed in acetone, removing the unexposed, uncross-linked material from below the grid through the holes in the metal. The structure is rinsed by immersion in iso-propyl alcohol. After drying the structure appears as shown in fig. \ref{fig:InGridSEM}. Our present detector has 25~$\mu$m diameter holes at a pitch of 55~$\mu$m; the multiplication gap (grid-to-anode) is $\sim$80~$\mu$m.

The chip is bonded to a readout board, to operate the Timepix chip we further need a small USB connected readout module and a PC with dedicated software \cite{USBreadout}. The readout board has an integrated gas chamber of app. \mbox{$10 \times 10 \times 1.5 \textrm{ cm}^3$}. Inside this volume an additional steel cathode mesh is mounted 5--10 mm above the grid. Above this mesh, centered on the location of the chip, is a UV transparent Suprasil quartz window.

After mounting the chip can be coated with CsI. For this the whole assembly without the top cover, window and cathode mesh is placed inside the evaporation chamber. The deposition is done by thermal evaporation in vacuum from a solid CsI source using resistive heating \cite{CsIBreskin}. In some cases the CsI layer is patterned using a simple Al foil shadow-mask during deposition. The typical CsI thickness is 220 nm.

\subsection{Measurement methods}
\begin{figure}[hbt]
\centering
\includegraphics[width=0.45\textwidth,keepaspectratio]{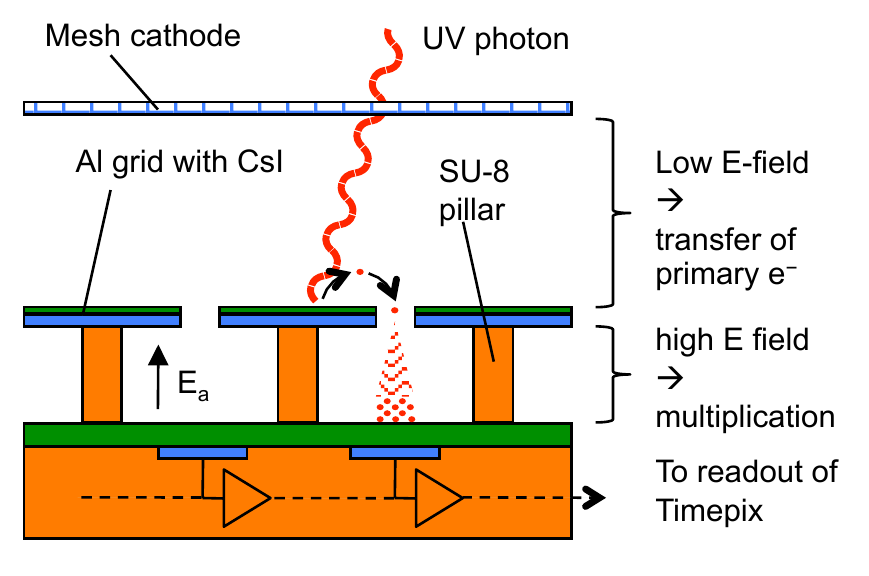}
\caption{Illustration of the device and its operation. Light enters through a window from the top; photoelectrons from a CsI photocathode on the grid are focused into the multiplication gap;  the resulting avalanche induces charge onto the CMOS pixel array of the Timepix chip Ð via a discharge-protecting film. Operation was in gas flow mode at 1~atm.}
\label{fig:deviceoperation}
\end{figure}

The different elements of the system and the detector operation principle are shown in fig. \ref{fig:deviceoperation}. Light impinges onto the photocathode layer on the top surface of the grid. The chip surface (anode) is grounded. The cathode mesh and the grid are typically biased at the same negative potential to prevent electrons to drift upwards. The electrons follow the evanescent field from the high field region below the grid. After drifting through the holes the charge is multiplied towards the anode. The charge cloud then falls onto one of the pixels of the matrix.

Charge signals from the detector are recorded from the grid, using a charge preamplifier, a shaping amplifier and a multi channel analyzer (MCA). The amplifier chain is calibrated by injecting a known charge from a test capacitor. The images are recorded with the Timepix chip using the readout methods discussed above.

\begin{figure}[hbt]
\centering
\includegraphics[width=0.45\textwidth,keepaspectratio]{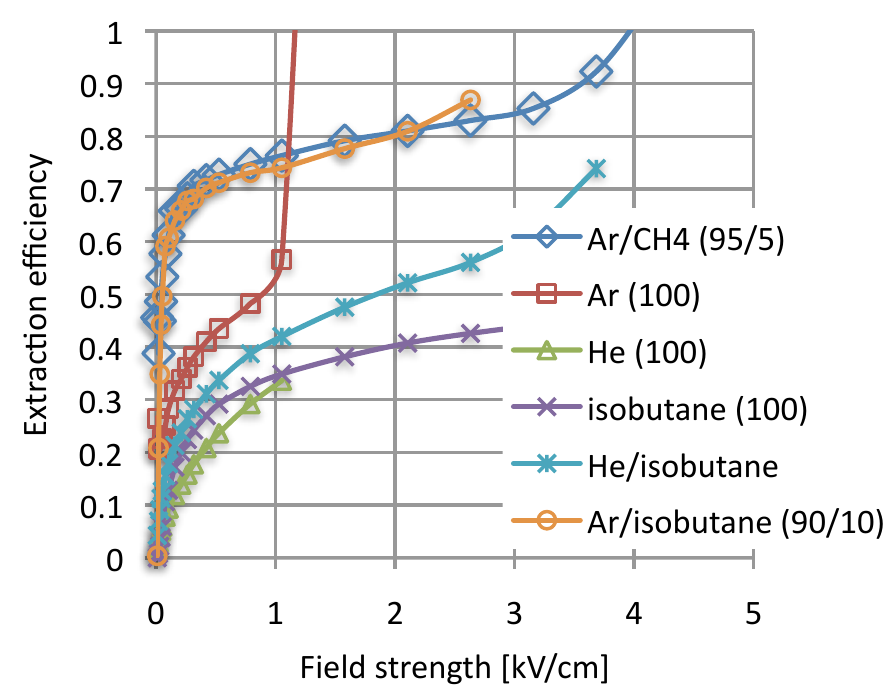}
\caption{Extraction efficiency of photoelectrons from a CsI photocathode into various gas mixtures (reference to vacuum). All gases were maintained at 1~atm.}
\label{fig:extraction}
\end{figure}

The devices can be operated with different gas mixtures; best results were obtained with a He/isobutane (80/20) mixture. Device operation is very stable in He/isobutane, several detectors have been operated for up to several weeks without any signs of degradation. Lower gains were reached in Ar based mixtures.
There are two drawbacks related to the use of He and isobutane. First, isobutane absorbs UV radiation. For our setup this effect is limited because of the low isobutane concentration and the small thickness of the gas layer ($\sim$1~cm). Secondly, in He there is rather high backscattering of photoelectrons on gas molecules compared to other noble gases \cite{coelho}. This reduces the effective quantum efficiency of the detector, even when mixed with isobutane. Fig. \ref{fig:extraction} shows measured extraction efficiency into various gases; the efficiency is the ratio between the extraction levels in gas and in vacuum, at the same bias and irradiation conditions. In He/isobutane (80/20) we reach an extraction level of around 50\% for normal operating conditions \mbox{($\sim 2$ kV/cm)}. For backscattering in other gas mixtures see \cite{Nemeas}.

\section{Results and discussion}

\subsection{Gain as determined from pulse measurements}
\begin{figure}[hbt]
\centering
\includegraphics[width=0.45\textwidth,keepaspectratio]{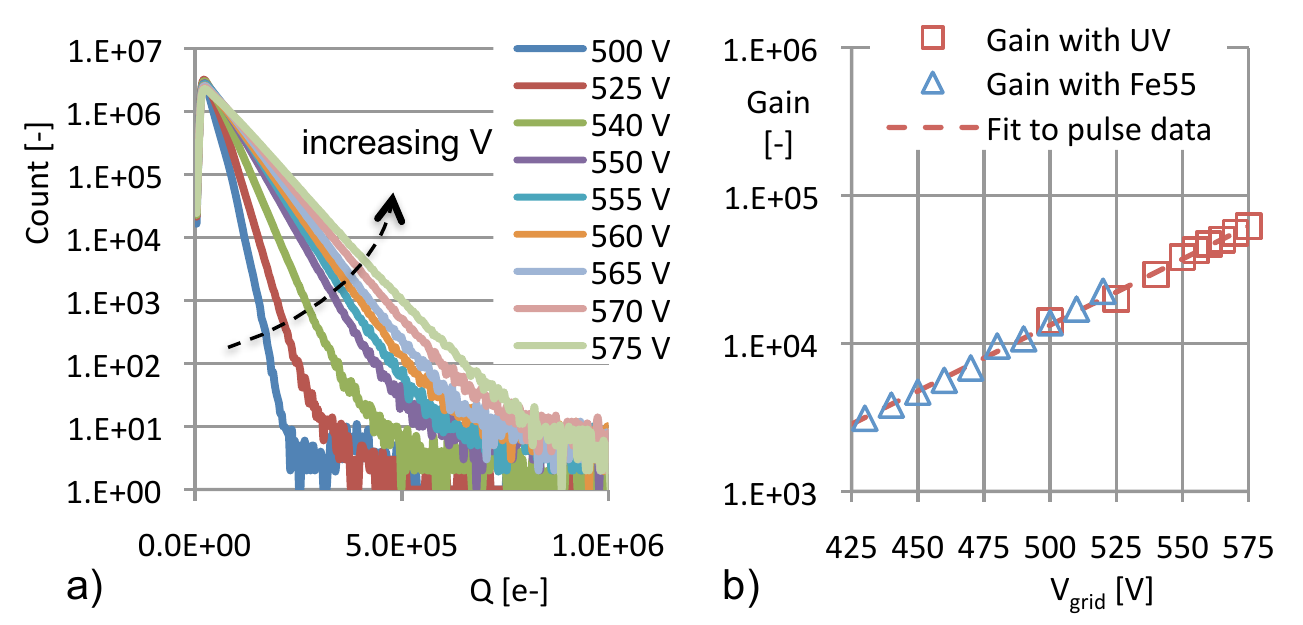}
\caption{a) Pulse-height spectra for a range of bias voltages, recorded on the grid under UV irradiation, b) Gain vs. bias voltage derived from pulse-height spectra of UV photons and of $^{55}$Fe 5.9~keV X-rays.}
\label{fig:pulsegain}
\end{figure}

The gain of the detector was determined by irradiating it with two different sources (UV light and $^{55}$Fe). In both cases the grid and the cathode mesh were biased at equal potentials and the anode was grounded; the measurements were done in He/isobutane (80/20) at atmospheric pressure. In one experiment we used an Ar(Hg) lamp with a pinhole aperture placed far from the detector. The light flux was further reduced with absorbers. The response of the detector to this irradiation was measured with the pre-amplifier attached to the grid. The exponential pulse-height spectra (typical for single-photon gas avalanche detectors) for a range of bias voltages are shown in fig. \ref{fig:pulsegain}a. The distributions are accurately described by eq. \ref{eq:distribution}, where \textit{N} is the amount of hits, \textit{Q} is charge, \textit{G} is the average gain and \textit{C} is a constant.

\begin{equation}
N(Q) = C \cdot 1/G \cdot \exp(-Q/G)
\label{eq:distribution}
\end{equation}

By making fits to the measured spectra we determined average gain \textit{G} for all voltages. We have also determined gain by measuring the average charge cloud arriving after multiplication of the initial charge from conversion of $^{55}$Fe photons (irradiating through a thin Kapton window) in the gas. The results of both gain measurements are plotted in fig. \ref{fig:pulsegain}b. There is good agreement between the two methods; with $^{55}$Fe we obtained a slope of 100~V/decade and with UV photons it is 110~V/decade. The maximum gain that was reached is $6\cdot10^4$.

\subsection{Spatial resolution using the slanted edge method}
\begin{figure}[hbt]
\centering
\includegraphics[width=0.45\textwidth,keepaspectratio]{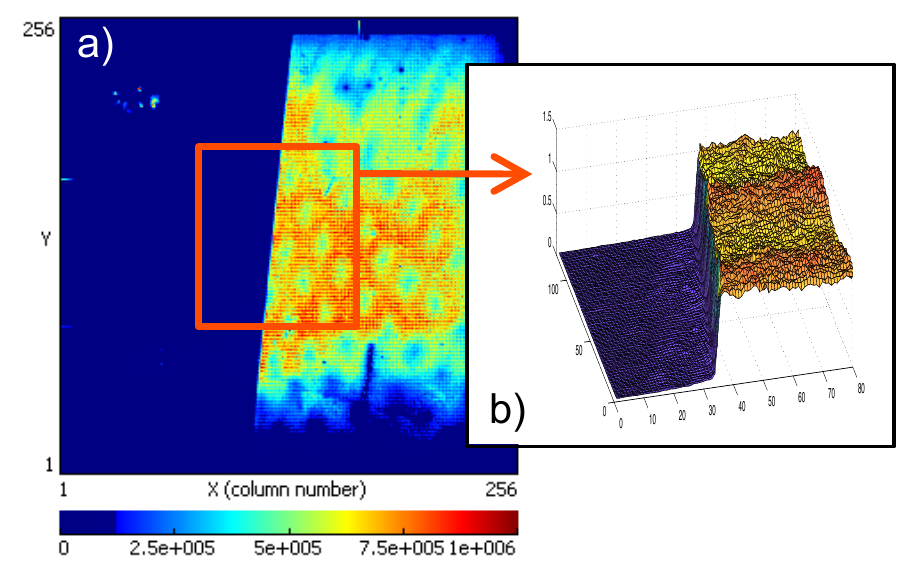}
\caption{a) An image of a slanted metal sheet, the data of a certain ROI is considered, the raw image data is corrected with an open frame measurement taken under the same conditions --- providing the step response on the right (b).}
\label{fig:step}
\end{figure}

\begin{figure}[hbt]
\centering
\includegraphics[width=0.35\textwidth,keepaspectratio]{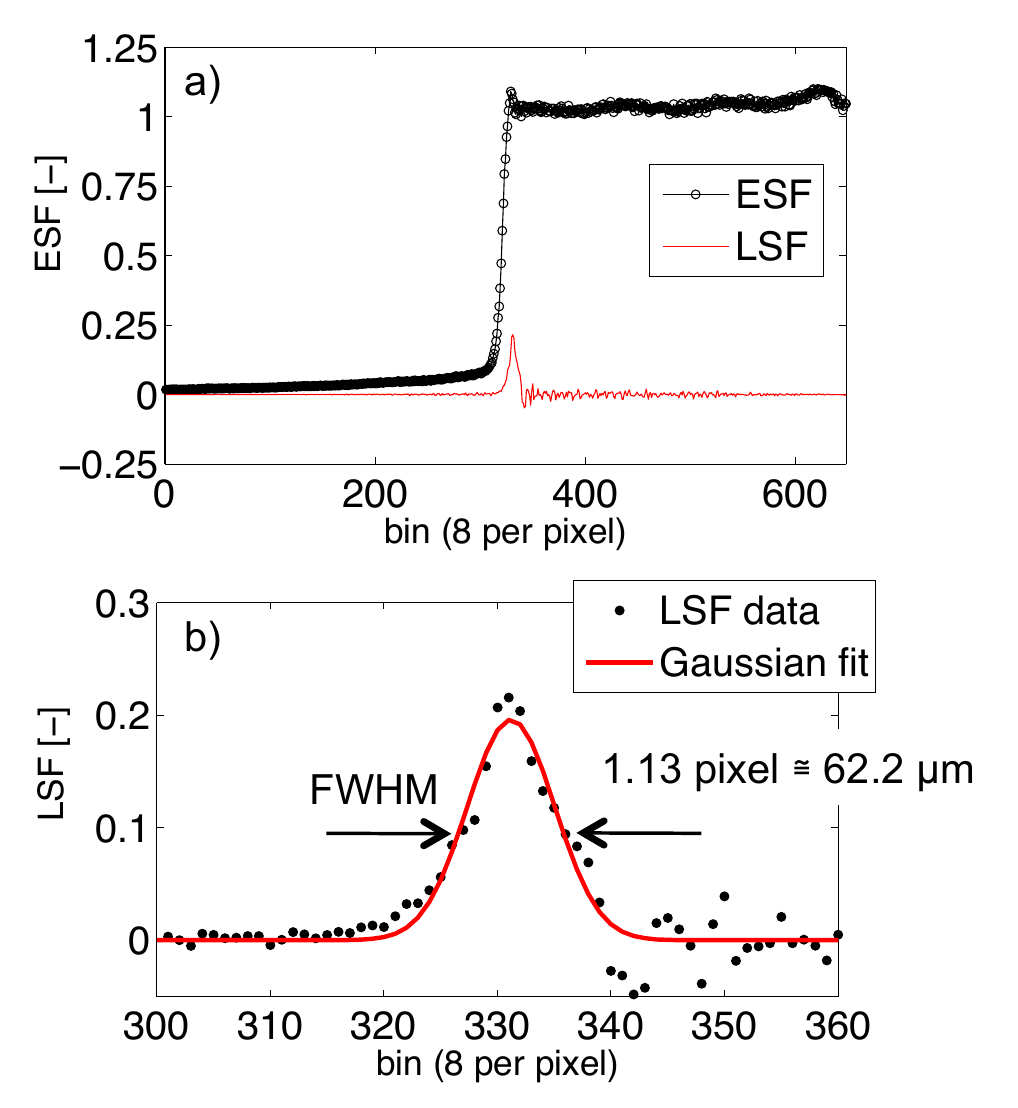}
\caption{a) ESF as derived from the rebinned data within the ROI, b) LSF data and Gaussian fit.}
\label{fig:LSF}
\end{figure}

To determine the spatial resolution we followed the ISO 12233 procedure \cite{slantededge}. Images were recorded under UV irradiation of the edge of a thin sheet of steel placed at a slight angle with respect to the pixel orientation. The response was corrected for fixed patterns by division with data of an open frame (flat field) measurement taken under the same conditions. The step response was analyzed within a certain Region Of Interest (ROI), as shown in fig. \ref{fig:step}a. For each line of pixels we took the derivative of the pixel counter value and determined the centroid; this defines the edge location within each line of pixels. A linear fit was performed through the centroid locations to derive a description of the edge. The data of all lines was rebinned (in this case with 8 bins per pixel) and simultaneously shifted according to the slant that was determined before. In this way we could combine all data in the ROI in one Edge Step Function (ESF). The derivative of the ESF is the Line Spread Function (LSF). A Gaussian curve was fitted to the LSF; the $1\sigma$ spread is \mbox{0.48 pixels} or \mbox{26.4 $\mu$m} and the FWHM is \mbox{1.13 pixels} or \mbox{62.2 $\mu$m}. The results are shown in fig. \ref{fig:LSF}.

By performing a Fourier transform of the LSF we obtained the Modulation Transfer Function (MTF) of the entire system. The MTF reaches 50\% at \mbox{0.4 lp/pixels} (\mbox{$\simeq$ 7 lp/mm}). The resolution limit is below the pixel pitch, at the Nyquist frequency \mbox{MTF = 0.32}. The resolution limit is reached at \mbox{0.8 lp/pixels} (\mbox{$\simeq$ 14 lp/mm}).

\subsection{Imaging results}
\begin{figure}[hbt]
\centering
\includegraphics[width=0.5\textwidth,keepaspectratio]{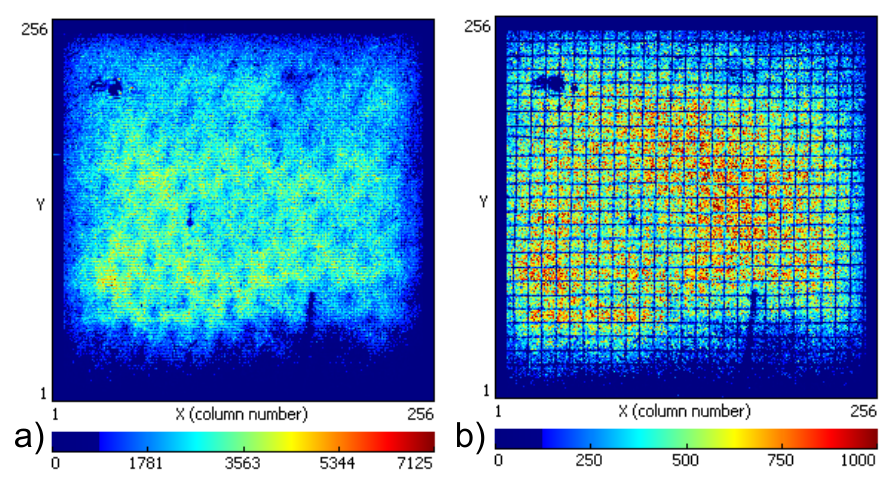}
\caption{2D images of flat-field UV irradiation of the detector shown in fig.~\ref{fig:deviceoperation} using He/isobutane (80/20) at 1 atm. a) Moir\'e pattern resulting from using a \mbox{56 $\mu$m} mesh above the 55~$\mu$m pitch detector and b) direct imaging of the mesh when the mesh pitch is increased to \mbox{500 $\mu$m}.}
\label{fig:mesh}
\end{figure}

The cathode mesh shown in fig. \ref{fig:deviceoperation}, needed to shape the field above the grid, modulates the light reaching the photocathode. In \mbox{fig. \ref{fig:mesh}a} we see a Moir\'e pattern which is the result of using a cathode mesh with a wire pitch of \mbox{56 $\mu$m} and a detector pitch (both grid and read-out pads) of \mbox{55 $\mu$m}. Using a coarser mesh, with a pitch of \mbox{500 $\mu$m}, the mesh wires are directly imaged by the detector (\mbox{fig. \ref{fig:mesh}b}). The high resolution permits observing mesh defects.

The images in fig. \ref{fig:mesh} were taken with a partial CsI coating. For experimental reasons not discussed here the bottom 15\% of the matrix was not coated with CsI, which naturally resulted in a lower response in that region.

\begin{figure}[hbt]
\centering
\includegraphics[width=0.5\textwidth,keepaspectratio]{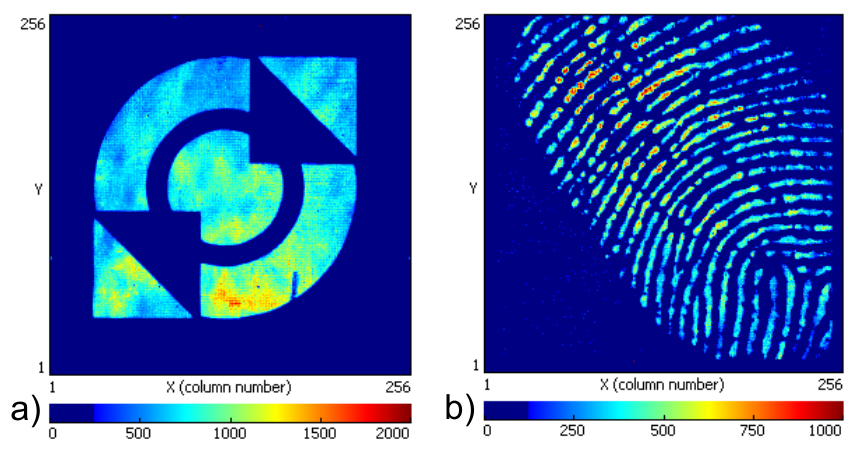}
\caption{a) Image of the logo of the University of Twente and b) image of a fingerprint left on the detector window.}
\label{fig:fingerprint}
\end{figure}

Fig. \ref{fig:fingerprint} shows two images obtained in He/isobutane (80/20) using TOT counting under UV-photon irradiation. Fig. \ref{fig:fingerprint}a is an image recorded with a steel mask of the logo of the University of Twente placed in front of the detector; fig. \ref{fig:fingerprint}b is an image of a fingerprint intentionally left on the detector window. The latter was obtained by correcting the raw image with that of an open frame reference image recorded under the same conditions. The fingerprint image was recorded without CsI on the Al grid. This shows that, in spite of the (very) low quantum efficiency, even without CsI the detector performs as a UV imaging device.

\section{Conclusions}
We have presented a monolithic gaseous UV-photon imaging detector with a CsI photocathode. The photocathode deposition and its operation are found to be adequate in combination with both the Micromegas multiplier as well as the imaging chip. Most experiments have been performed in He/isobutane (80/20), in which the extraction efficiency is around 50\%. The maximum gain that was reached is $6\cdot10^4$.

Imaging experiments indicate good spatial resolution, the LSF has a spread of 26.4 $\mu$m, MTF50 is reached at 0.4~lp/pixel ($\simeq$ 7~lp/mm). High quality 2D images were recorded.

These encouraging results suggest further investigations directed towards the search for operation conditions and gas-mixtures assuring better photoelectron collection efficiencies (lower backscattering); investigations of absolute photon detection efficiencies and of ion-blocking (ion-feedback) are foreseen.
The new device could find applications in high-resolution imaging of UV  photons. Furthermore, being made of UHV-compatible materials, it could be applicable as a multiplication and readout element also in cascaded visible-light imaging gas photomultipliers \cite{Lyashenko}.

\section*{Acknowledgements}
The authors acknowledge Sander Smits (of the University of Twente), Max Chefdeville, Fred Hartjes, Joop R\"ovekamp and Jan Timmermans (all with NIKHEF) and Moshe Klin and Alexey Lyashenko (both from the Weizmann Institute) for their help during the manufacturing of the test devices and the measurements. This research is funded by Dutch technology foundation STW through project TET-6630. A. Breskin is the W.P. Reuther Professor of Research in The Peaceful Use of Atomic Energy.

\end{document}